\documentclass[article, onecolumn, 11pt, superscriptaddress]{revtex4-2}
\usepackage{eurosym}
\usepackage[utf8]{inputenc}
\usepackage{amsmath}
\usepackage{amsfonts}
\usepackage{amssymb}
\usepackage{hyperref}
\usepackage{graphicx}
\usepackage{color}

\begin{document}
\title{Phase Transitions and Order Parameters in Correlation Matrices: A
Wishart-Ensemble Perspective on the Largest Eigenvalue}

\author{Roberto da Silva}
\affiliation{Instituto de F\'{\i}sica, Universidade Federal do Rio Grande do Sul (UFRGS), Porto Alegre, RS 91501-970, Brazil}

\author{Antonio Mihara}
\affiliation{Departamento de F\'{\i}sica, Universidade Federal de S\~ao Paulo (UNIFESP), Diadema, SP 09913-030, Brazil}

\author{Henrique Tramontina}
\affiliation{Instituto de F\'{\i}sica, Universidade Federal do Rio Grande do Sul (UFRGS), Porto Alegre, RS 91501-970, Brazil}

\author{Sandra D. Prado}
\affiliation{Instituto de F\'{\i}sica, Universidade Federal do Rio Grande do Sul (UFRGS), Porto Alegre, RS 91501-970, Brazil}

\date{\today}

\begin{abstract}

We investigate the properties of the largest eigenvalue of correlation
matrices within the framework of Wishart ensembles. In this work, we propose
the largest eigenvalue as an effective empirical order parameter for detecting phase
transitions in chaotic and spin systems, drawing an analogy between its derivatives and thermodynamic response functions derived from the free energy, however not necessarily linked to a critical divergence originally observed in the context of phase transitions theory.
\end{abstract}

\maketitle

\section{Introduction}

Eigenvalues translate the internal organization of a matrix into spectral
form. In this translation, the largest eigenvalue plays a particularly
prominent role: it represents the most amplified mode allowed by the matrix
structure. This interpretation is deeply connected with principal component
analysis, where the leading eigenvalue of a covariance or correlation matrix
measures the variance captured by the first principal component \cite%
{Pearson1901,Hotelling1933,Jolliffe2002,Jolliffe2016}. For matrices built
from physical, statistical, or dynamical data, this leading mode often
corresponds to the strongest collective component present in the system.
Thus, the largest eigenvalue provides a natural way to pass from many
microscopic relations to a single macroscopic indicator of organization.

This idea is especially powerful when the matrix entries encode
correlations. A system composed of many weakly related degrees of freedom
may exhibit a spectrum with no dominant component, reflecting the absence of
a coherent collective direction. By contrast, when many variables begin to
fluctuate in a coordinated way, this coherence is accumulated spectrally and
may produce a leading eigenvalue that stands out from the remaining
spectrum. This mechanism has been extensively explored in empirical
correlation matrices, particularly in financial systems, where the largest
eigenvalue is commonly associated with a collective market mode, while the
remaining eigenvalues contain sectorial or noisy components \cite%
{Plerou2002,Bouchaud2011,Bun2017}.

The distinction between a leading eigenvalue embedded in the spectral bulk
and one separated from it is central to modern spectral analysis. Random
matrix theory \cite{Mehta2004} provides a natural reference framework for this distinction.
The Wishart ensemble and the Marchenko--Pastur law describe the spectral
distribution expected from large random covariance or correlation matrices
in the absence of genuine structure \cite%
{Wishart1928,MarchenkoPastur1967}. Within this framework, the
largest eigenvalue plays a special role because it lies at the spectral
edge, where universal fluctuation laws such as the Tracy--Widom distribution
arise \cite{TracyWidom1994,Johnstone2001}. Therefore, a leading eigenvalue
that remains compatible with the random-matrix edge may be interpreted as a
finite-size or noise-induced fluctuation, whereas a systematic separation
from the bulk indicates the presence of an organized component not explained
by randomness alone.

This spectral separation is also at the heart of spiked random matrix
models, in which a low-rank structured signal is added to a noisy
background. In such models, the largest eigenvalue undergoes a transition:
below a critical signal strength it remains absorbed by the random bulk,
while above it detaches and becomes an outlier carrying information about
the underlying signal \cite{Baik2005,BaikSilverstein2006}. This provides a
precise mathematical paradigm for interpreting large leading eigenvalues as
signatures of collective or coherent structure.

For this reason, largest-eigenvalue methods are naturally connected to the
characterization of phase transitions. In physical systems, transitions are
accompanied by changes in the way degrees of freedom correlate and organize.
These changes may be reflected in the spectral properties of matrices
constructed from configurations, time series, correlation functions,
adjacency structures, or dynamical observables. In networked dynamical
systems, for example, the onset of synchronization and other collective
processes can be controlled by the largest eigenvalue of the adjacency
matrix \cite{Restrepo2005,Restrepo2007}. More generally, the growth,
fluctuation, or separation of the largest eigenvalue can be interpreted as a
spectral signature of the emergence of collective behavior, making it a
useful tool for detecting and characterizing critical phenomena.

Random Matrix Theory (RMT) has proved to be an indispensable tool for
understanding the statistical properties of complex systems, ranging
from nuclear physics to financial markets and neural networks \cite%
{Mehta2004, Akemann2011}. Among the various ensembles, the Wishart ensemble, which describes sample covariance and correlation
matrices, plays a central
role in identifying signal from noise in high-dimensional data \cite%
{Wishart1928}.

In recent years, the behavior of the extreme eigenvalues, particularly the
largest eigenvalue $\lambda _{max}$, has attracted significant interest. For
a standard Wishart matrix $W=\frac{1}{T}XX^{\dagger }$, where $X$ is an $%
N\times T$ matrix of independent Gaussian variables, the distribution of $%
\lambda _{max}$ is known to follow the Tracy-Widom distribution in the limit
of large $N$ and $T$ \cite{TracyWidom1994,Johnstone2001}. However, when the system deviates from pure randomness, for example in the
presence of strong correlations or structural transitions, the largest
eigenvalue often separates of the bulk spectrum (the Marchenko-Pastur distribution), signaling the
emergence of a macroscopic order \cite{Baik2005}.

In the context of physical systems, such as spin glasses or dynamical
systems exhibiting chaotic transitions, the largest eigenvalue can be
interpreted as a measure of global coherence. Analogously to magnetization in ferromagnetic systems, $\lambda_{max}$ acts as an indicator
of the collective behavior of the underlying degrees of freedom. By examining the mean value, fluctuations, and distributional shape of
$\lambda_{max}$, one can determine whether changes in the underlying
correlation structure provide spectral signatures of a transition.

In this paper, we explore the use of the largest eigenvalue of correlation
matrices as an empirical order parameter to characterize phase transitions in
chaotic maps and spin systems. We show that the first derivative of $\lambda _{max}$ with respect to the
relevant control parameter exhibits behavior analogous, at an empirical
level, to thermodynamic response functions derived from the free energy. Depending on the system, we observe either a sharp variation or a
pronounced extremum near the expected critical parameter. By applying this framework to various models, we provide
a robust metric for identifying the onset of order in complex dynamical
regimes. As test cases, we estimate the critical temperature of the two-dimensional
Ising model and characterize the transition to chaos in the logistic map,
while also monitoring the distribution of the largest eigenvalue. In the next section, we discuss the methodology and establish the connection
between the largest eigenvalue and random correlation matrices within the
Wishart-ensemble framework. In section \ref{Sec:Results} we present our results. Finally, in Section \ref{Sec:Conclusions}, we summarize our main findings
and present the conclusions.

\section{Maximal Eigenvalue and Correlation Random Matrices}

\label{Sec:Maximal_Eigenvalue}

Methods based on the largest eigenvalue provide a natural way to identify
collective organization in complex systems. Let $C$ be a real symmetric
matrix, for instance a correlation matrix. By the Rayleigh--Ritz variational
principle, its largest eigenvalue is given by 
\begin{equation}
\lambda_{\max}=\max_{\Vert v\Vert=1}v^{T}Cv .  \label{eq:rayleigh_lmax}
\end{equation}
When $C$ is a correlation matrix associated with a set of standardized
variables $\{x_i\}_{i=1}^{N}$, the quadratic form in Eq. (\ref%
{eq:rayleigh_lmax}) has a direct statistical meaning. Defining the
collective variable 
\begin{equation}
X_v=\sum_{i=1}^{N}v_i x_i ,  \label{eq:collective_variable}
\end{equation}
one obtains 
\begin{equation}
\mathrm{Var}(X_v)=\mathrm{Var}\left(\sum_{i=1}^{N}v_i x_i\right).
\label{eq:variance_definition}
\end{equation}
Since the variables are standardized and $C_{ij}$ denotes their correlation
matrix, we have 
\begin{equation}
C_{ij}=\left\langle x_i x_j\right\rangle .  \label{eq:corr_matrix_definition}
\end{equation}
Therefore, 
\begin{equation}
\mathrm{Var}(X_v) = \left\langle \left(\sum_{i=1}^{N}v_i x_i\right)
\left(\sum_{j=1}^{N}v_j x_j\right) \right\rangle .
\label{eq:variance_expansion_1}
\end{equation}
Expanding the sums gives 
\begin{equation}
\mathrm{Var}(X_v) = \sum_{i,j=1}^{N}v_i v_j \left\langle x_i
x_j\right\rangle .  \label{eq:variance_expansion_2}
\end{equation}
Using Eq. (\ref{eq:corr_matrix_definition}), this becomes 
\begin{equation}
\mathrm{Var}(X_v)=\sum_{i,j=1}^{N}v_i C_{ij}v_j .
\label{eq:variance_expansion_3}
\end{equation}
In matrix notation, we finally obtain 
\begin{equation}
\mathrm{Var}(X_v)=v^{T}Cv .  \label{eq:variance_collective_mode}
\end{equation}
Combining Eqs. (\ref{eq:rayleigh_lmax}) and (\ref%
{eq:variance_collective_mode}), the largest eigenvalue can be interpreted as
the maximum variance attainable by a normalized collective mode, 
\begin{equation}
\lambda_{\max}=\max_{\Vert v\Vert=1}\mathrm{Var}(X_v).
\label{eq:lmax_max_variance}
\end{equation}
In this sense, $\lambda_{\max}$ is not merely a spectral observable. It
quantifies the strength of the dominant collective fluctuation present in
the system. The associated eigenvector identifies the linear combination of
variables that realizes this maximum, while the corresponding eigenvalue
measures the intensity of the collective mode.

This interpretation can be made explicit through a simple illustrative
calculation. Consider an idealized correlation matrix in which all
off-diagonal correlations are equal to a common value $\rho$. This matrix
can be written as 
\begin{equation}
C_{ij}=(1-\rho)\delta_{ij}+\rho .  \label{eq:constant_corr_matrix}
\end{equation}
If $i=j$, Eq. (\ref{eq:constant_corr_matrix}) gives $C_{ii}=1$. If $i\neq j$%
, it gives $C_{ij}=\rho$. Equivalently, in matrix form, 
\begin{equation}
C=(1-\rho)I+\rho J ,  \label{eq:corr_matrix_decomposition}
\end{equation}
where $I$ is the identity matrix and $J$ is the matrix whose entries are all
equal to one. The uniform vector is 
\begin{equation}
u={\frac{1}{\sqrt{N}}}(1,1,\ldots,1)^{T}.  \label{eq:uniform_vector}
\end{equation}
Since each row of $J$ contains $N$ entries equal to one, this vector
satisfies 
\begin{equation}
Ju=Nu .  \label{eq:J_uniform_eigenvalue}
\end{equation}
Thus, the uniform vector is an eigenvector of $J$ with eigenvalue $N$. All
vectors orthogonal to $u$ are associated with zero eigenvalue of $J$.
Therefore, the spectrum of $J$ is composed of one eigenvalue equal to $N$
and $N-1$ eigenvalues equal to zero.

Using Eq. (\ref{eq:corr_matrix_decomposition}), the eigenvalue associated
with the uniform mode is  
\begin{equation}
\lambda_1=1+(N-1)\rho .  \label{eq:first_eigenvalue}
\end{equation}
For the remaining $N-1$ directions, orthogonal to the uniform vector, the
corresponding eigenvalues are 
\begin{equation}
\lambda_2=\lambda_3=\cdots=\lambda_N=1-\rho .
\label{eq:remaining_eigenvalues}
\end{equation}
Thus, for $\rho>0$, the largest eigenvalue is 
\begin{equation}
\lambda_{\max}=1+(N-1)\rho .  \label{eq:lmax_constant_corr}
\end{equation}
This expression shows that even weak correlations can generate a large
eigenvalue when they are coherently distributed over the whole system. In
particular, for large $N$, one has 
\begin{equation}
\lambda_{\max}\simeq N\rho .  \label{eq:lmax_scaling}
\end{equation}
Therefore, if the average correlation remains finite as $N$ increases, the
largest eigenvalue becomes a macroscopic quantity. This simple result
explains why $\lambda_{\max}$ is highly sensitive to collective
organization, ordering phenomena, synchronization, and critical fluctuations.

Random Matrix Theory provides the natural null model against which such
collective effects can be tested. Consider a data matrix $X$ of dimension $%
N\times T$, where $N$ is the number of variables and $T$ is the number of
observations. If the entries of $X$ are independent, centered, and have unit
variance, the empirical covariance matrix can be written as 
\begin{equation}
C={\frac{1}{T}}XX^{T}.  \label{eq:wishart_matrix_definition}
\end{equation}
This construction defines the Wishart ensemble \cite{Wishart1928}. In this
null model, the variables contain no genuine correlations, and any
nontrivial spectral structure arises only from finite-size fluctuations.

Let $\lambda_1,\lambda_2,\ldots,\lambda_N$ be the eigenvalues of $C$. For a
real Wishart ensemble, the joint probability density of the eigenvalues can
be written, apart from a normalization constant, as 
\begin{equation}
P(\lambda_1,\ldots,\lambda_N) = {\frac{1}{Z_{N,T}^{(\beta)}}}A_1A_2A_3 .
\label{eq:wishart_jpdf_1}
\end{equation}
The factor $A_1$ is 
\begin{equation}
A_1= \exp\left[-{\frac{\beta T}{2}}\sum_{i=1}^{N}\lambda_i\right].
\label{eq:wishart_jpdf_2}
\end{equation}
The factor $A_2$ is 
\begin{equation}
A_2= \prod_{i=1}^{N} \lambda_i^{\beta(T-N+1)/2-1}.  \label{eq:wishart_jpdf_3}
\end{equation}
The factor $A_3$ is the Vandermonde contribution, 
\begin{equation}
A_3= \prod_{i<j}|\lambda_i-\lambda_j|^{\beta}.  \label{eq:wishart_jpdf_4}
\end{equation}
The parameter $\beta$ distinguishes the symmetry class of the ensemble: $%
\beta=1$ for real matrices and $\beta=2$ for complex matrices. The
Vandermonde factor in Eq. (\ref{eq:wishart_jpdf_4}) represents eigenvalue
repulsion, one of the central features of random matrix spectra \cite%
{Mehta2004,Akemann2011}.

The average density of eigenvalues is obtained by integrating the joint
density over all eigenvalues except one. The finite-size spectral density is
defined as 
\begin{equation}
\rho_N(\lambda) = {\frac{1}{N}} \left\langle
\sum_{i=1}^{N}\delta(\lambda-\lambda_i) \right\rangle .
\label{eq:finite_density_definition}
\end{equation}
Equivalently, in terms of the joint probability density, 
\begin{equation}
\rho_N(\lambda) = \int_{0}^{\infty}d\lambda_2\cdots
\int_{0}^{\infty}d\lambda_N P(\lambda,\lambda_2,\ldots,\lambda_N).
\label{eq:one_point_integral}
\end{equation}
In the asymptotic limit $N,T\rightarrow \infty$, with the ratio 
\begin{equation}
q={\frac{N}{T}}  \label{eq:q_ratio}
\end{equation}
kept fixed and $0<q\leq 1$, this density converges to the Marchenko--Pastur
distribution \cite{MarchenkoPastur1967}. The Marchenko--Pastur density is
given by 
\begin{equation}
\rho_{\mathrm{MP}}(\lambda) = {\frac{1}{2\pi q\lambda}} \sqrt{%
(\lambda_+-\lambda)(\lambda-\lambda_-)} ,  \label{eq:mp_density}
\end{equation}
for $\lambda_-\leq \lambda\leq \lambda_+$, and vanishes outside this
interval. The spectral edges are 
\begin{equation}
\lambda_-=(1-\sqrt{q})^2 ,  \label{eq:mp_lower_edge}
\end{equation}
and 
\begin{equation}
\lambda_+=(1+\sqrt{q})^2 .  \label{eq:mp_upper_edge}
\end{equation}
For variables with variance $\sigma^2$, the edges become 
\begin{equation}
\lambda_- = \sigma^2(1-\sqrt{q})^2  \label{eq:mp_lower_edge_sigma}
\end{equation}
and 
\begin{equation}
\lambda_+ = \sigma^2(1+\sqrt{q})^2 .  \label{eq:mp_upper_edge_sigma}
\end{equation}
For a correlation matrix constructed from standardized variables, the mean
eigenvalue is equal to one. Indeed, 
\begin{equation}
{\frac{1}{N}}\sum_{i=1}^{N}\lambda_i = {\frac{1}{N}}\mathrm{Tr}\,C .
\label{eq:mean_eigenvalue_trace}
\end{equation}
Since $C_{ii}=1$, one obtains 
\begin{equation}
{\frac{1}{N}}\mathrm{Tr}\,C=1 .  \label{eq:mean_eigenvalue_corr}
\end{equation}
Thus, the Marchenko--Pastur law gives the spectral distribution expected
from purely random correlations. Eigenvalues inside the interval $%
[\lambda_-,\lambda_+]$ are compatible with finite-size noise, whereas
eigenvalues systematically outside this interval reveal structure not
contained in the null Wishart model.

The largest eigenvalue is the extreme spectral variable 
\begin{equation}
\lambda_{\max}=\max\{\lambda_1,\lambda_2,\ldots,\lambda_N\}.
\label{eq:lmax_definition}
\end{equation}
Its cumulative distribution is obtained from the joint density by requiring
that all eigenvalues lie below a threshold $s$, 
\begin{equation}
\mathrm{Prob}(\lambda_{\max}\leq s) = \int_{0}^{s}d\lambda_1
\int_{0}^{s}d\lambda_2 \cdots \int_{0}^{s}d\lambda_N
P(\lambda_1,\ldots,\lambda_N).  \label{eq:lmax_cdf_finite}
\end{equation}
The corresponding probability density is 
\begin{equation}
p_{\max}(s) = {\frac{d}{ds}}\mathrm{Prob}(\lambda_{\max}\leq s).
\label{eq:lmax_pdf_finite}
\end{equation}
In the Wishart null model, the largest eigenvalue is concentrated near the
upper Marchenko--Pastur edge $\lambda_+$. However, at finite $N$ it still
fluctuates around this edge. After appropriate centering and scaling, these
edge fluctuations converge to the Tracy--Widom distribution \cite%
{TracyWidom1994,Johnstone2001}.

For the real Wishart ensemble, this limiting result can be written as 
\begin{equation}
\mathrm{Prob}\left[ {\frac{\lambda_{\max}-\mu_{N,T}}{\sigma_{N,T}}} \leq s %
\right] \longrightarrow F_1(s),  \label{eq:tw_limit}
\end{equation}
where $F_1(s)$ is the Tracy--Widom distribution associated with $\beta=1$. A
commonly used finite-size centering is 
\begin{equation}
\mu_{N,T} = {\frac{1}{T}} (\sqrt{T-1/2}+\sqrt{N-1/2})^2 .
\label{eq:johnstone_mu}
\end{equation}
The corresponding scaling is 
\begin{equation}
\sigma_{N,T} = {\frac{1}{T}} (\sqrt{T-1/2}+\sqrt{N-1/2})B_{N,T}^{1/3},
\label{eq:johnstone_sigma_1}
\end{equation}
where 
\begin{equation}
B_{N,T} = {\frac{1}{\sqrt{T-1/2}}} + {\frac{1}{\sqrt{N-1/2}}} .
\label{eq:johnstone_sigma_2}
\end{equation}
Equations (\ref{eq:tw_limit})--(\ref{eq:johnstone_sigma_2}) express the fact
that the largest eigenvalue of a purely random correlation matrix has a
universal limiting distribution at the spectral edge. Therefore, in the
absence of true correlations, $\lambda_{\max}$ is expected to remain close
to $\lambda_+$, with fluctuations described by Tracy--Widom statistics.

The importance of this result is both conceptual and practical. The
Marchenko--Pastur distribution describes the random bulk of eigenvalues,
whereas the Tracy--Widom law describes the random fluctuations of the upper
edge. Together, they provide a criterion for separating noise from coherent
structure. If the observed largest eigenvalue is compatible with the
Marchenko--Pastur edge and with Tracy--Widom fluctuations, it can be
interpreted as part of the random background. If, on the other hand, $%
\lambda_{\max}$ separates from $\lambda_+$, this separation indicates that
the data contain a dominant component that cannot be explained by random
uncorrelated variables.

This mechanism is made explicit in spiked covariance models. In such models,
one assumes that the population covariance matrix contains a low-rank
structured perturbation superposed on an otherwise random background. If $%
\theta$ is a population eigenvalue larger than the noise background, then
the associated empirical eigenvalue remains absorbed by the random bulk when
the signal is weak. Above a critical strength, however, it detaches from the
Marchenko--Pastur support and becomes an outlier \cite%
{Baik2005,BaikSilverstein2006}. In the simplest spiked model, the separation
occurs when 
\begin{equation}
\theta>1+\sqrt{q}.  \label{eq:bbp_threshold}
\end{equation}
The corresponding sample eigenvalue is asymptotically located at 
\begin{equation}
\lambda_{\mathrm{out}} = \theta \left(1+{\frac{q}{\theta-1}}\right).
\label{eq:spiked_outlier}
\end{equation}
This transition of the largest eigenvalue from the edge of the random bulk
to an isolated outlier is the spectral analogue of the emergence of a
macroscopic coherent mode.

For physical systems, this picture suggests a direct interpretation. A
disordered or weakly correlated regime should produce correlation matrices
whose spectra are close to the Wishart prediction, with $\lambda_{\max}$
near the Marchenko--Pastur edge. As the control parameter of the system is
changed and collective correlations develop, the largest eigenvalue may
grow, fluctuate anomalously, or separate from the random bulk. In this
regime, $\lambda_{\max}$ behaves as a spectral indicator of organization.

This is the basis of the approach adopted in the present work. We construct
correlation matrices from dynamical or statistical data and monitor the
average largest eigenvalue as a function of the relevant control parameter.
If the system undergoes a transition, the change in its collective
correlation structure should be reflected in the behavior of $%
\left\langle\lambda_{\max}\right\rangle$. In particular, a sharp variation
of this quantity, or a peak in its derivative with respect to the control
parameter, can be interpreted as a spectral signature of the transition. In
this sense, the largest eigenvalue plays a role analogous to an empirical
order parameter: it condenses information about the collective organization
of many degrees of freedom into a single spectral observable.

\section{Results}

\label{Sec:Results}

To obtain our results, we construct a cross-correlation
matrix $\mathcal{C}$ by using time series iterations of the observable $O$
which, in this work, is either the time evolution of the magnetization per
spin in the two-dimensional Ising model, simulated with the Metropolis
algorithm \cite{Metropolis1953,newman1999monte}, or the iterates of the logistic map.

The elements of this matrix for a general observable $O$ can be defined as
follows:

\begin{equation*}
\mathcal{C}_{ij}\mathcal{=}\frac{\left\langle O^{(i)}O^{(j)}\right\rangle
-\left\langle O^{(i)}\right\rangle \left\langle O^{(j)}\right\rangle }{%
\sigma _{O^{(i)}}\sigma _{O^{(j)}}}
\end{equation*}%
These elements are calculated from two time evolutions of length $T$: $%
O_{0}^{(i)},O_{1}^{(i)},...,O_{T-1}^{(i)}$ and $%
O_{0}^{(j)},O_{1}^{(j)},...,O_{T-1}^{(j)}$, such that:%
\begin{equation*}
\left\langle O^{(k)}\right\rangle =\frac{1}{T}\sum_{p=0}^{T-1}O_{p}^{(k)}%
\text{, }\left\langle O^{(i)}O^{(j)}\right\rangle =\frac{1}{T}%
\sum_{p=0}^{T-1}O_{p}^{(i)}O_{p}^{(j)}\text{ }
\end{equation*}%
with $\sigma _{O^{(k)}}^{2}=\frac{1}{(T-1)}\sum_{p=0}^{T-1}(O_{p}^{(k)}-\left%
\langle O^{(k)}\right\rangle )^{2}\approx \left\langle
O^{(k)2}\right\rangle -\left\langle O^{(k)}\right\rangle ^{2}$.

It is important to note that the matrix $\mathcal{C}$, of dimension $N$, can
be obtained from the standardized time-evolution matrix\ $\mathcal{M}$: 
\begin{equation}
\mathcal{M}=\left( 
\begin{array}{cccc}
\frac{O_{0}^{(1)}-\left\langle O^{(1)}\right\rangle }{\sigma _{O^{(1)}}} & 
\frac{O_{0}^{(2)}-\left\langle O^{(2)}\right\rangle }{\sigma _{O^{(2)}}} & 
\cdots & \frac{O_{0}^{(N)}-\left\langle O^{(N)}\right\rangle }{\sigma
_{O^{(N)}}} \\ 
\frac{O_{1}^{(1)}-\left\langle O^{(1)}\right\rangle }{\sigma _{O^{(1)}}} & 
\frac{O_{1}^{(2)}-\left\langle O^{(2)}\right\rangle }{\sigma _{O^{(2)}}} & 
& \frac{O_{1}^{(N)}-\left\langle O^{(N)}\right\rangle }{\sigma _{O^{(N)}}}
\\ 
\vdots & \vdots &  & \vdots \\ 
\frac{O_{T-1}^{(1)}-\left\langle O^{(1)}\right\rangle }{\sigma _{O^{(1)}}} & 
\frac{O_{T-1}^{(2)}-\left\langle O^{(2)}\right\rangle }{\sigma _{O^{(2)}}} & 
& \frac{O_{T-1}^{(N)}-\left\langle O^{(N)}\right\rangle }{\sigma _{O^{(N)}}}%
\end{array}%
\right)
\end{equation}

This allows us to verify that: 
\begin{equation}
\mathcal{C=}\frac{1}{T}\mathcal{M}^{t}\mathcal{M}.
\label{Eq:Covariance_matrix}
\end{equation}

If $O_{0}^{(i)},O_{1}^{(i)},...,O_{T-1}^{(i)}$ form a set of independent
random variables, Eq. (\ref{Eq:Covariance_matrix}) defines a real Wishart
matrix, and the results discussed in the previous section apply.

We first explore this approach by introducing the magnetization matrix
element $O_{i-1}^{(j)}=m_{ij}$ representing the magnetization of the $j$th
time series at the $i$th Monte Carlo (MC) step for a system containing $L^{d}$ spins. Here, $i=1,...,T$, while $j=1,...,N$
labels the independent time evolutions used to construct the correlation
matrix. For simplicity, we adopt $d=2$, the lowest spatial dimension in which the
nearest-neighbor Ising model exhibits a finite-temperature phase transition \cite{Onsager1944}.

For the second application, we consider $O_{i-1}^{(j)}=x_{ij}$, representing the logistic map \cite{May1976}
value at the $i$th iteration of the $j$th evolution. For the Ising model, each evolution uses an independent sequence of random
numbers but starts from the same fully ordered ferromagnetic state
($m_{0}=1$). For the logistic map, each evolution
corresponds to a different initial condition. Thus, in both cases, we consider an ensemble of $N_{run}$ different matrices.

We simulate the Ising model for $T=300$ MC steps and use $N=100$
independent evolutions to construct each matrix. The procedure is repeated
for $N_{run}=1000$ matrices. We then compute the average largest
eigenvalue and its variance: 
\begin{equation}
\begin{array}{rrl}
\left\langle \lambda _{\max }\right\rangle & = & \frac{1}{N_{run}}%
\sum_{j=1}^{N_{run}}\lambda _{\max }^{(j)} \\ 
&  &  \\ 
\left\langle \left( \Delta \lambda_{\max} \right) ^{2}\right\rangle =\left\langle
\left( \lambda_{\max} -\left\langle\lambda _{\max }\right\rangle\right) ^{2}\right\rangle & \approx & \frac{1%
}{N_{run}}\sum_{j=1}^{N_{run}}\lambda _{\max }^{(j)\ 2}-\frac{1}{N_{run}^{2}}%
\left( \sum_{j=1}^{N_{run}}\lambda _{\max }^{(j)\ }\right) ^{2}%
\end{array}
\label{Eq:moments}
\end{equation}

We therefore monitor $\left\langle \lambda _{\max }\right\rangle $, $%
\left\langle \left( \Delta \lambda \right) ^{2}\right\rangle $ and $\frac{%
d\left\langle \lambda _{\max }\right\rangle }{dT}$ as functions of $\frac{%
k_{B}T}{J}$. The results are presented in Fig. \ref%
{Fig:phase_transition_ising}.

\begin{figure}[tbp]
\begin{center}
\includegraphics[width=0.8\columnwidth]{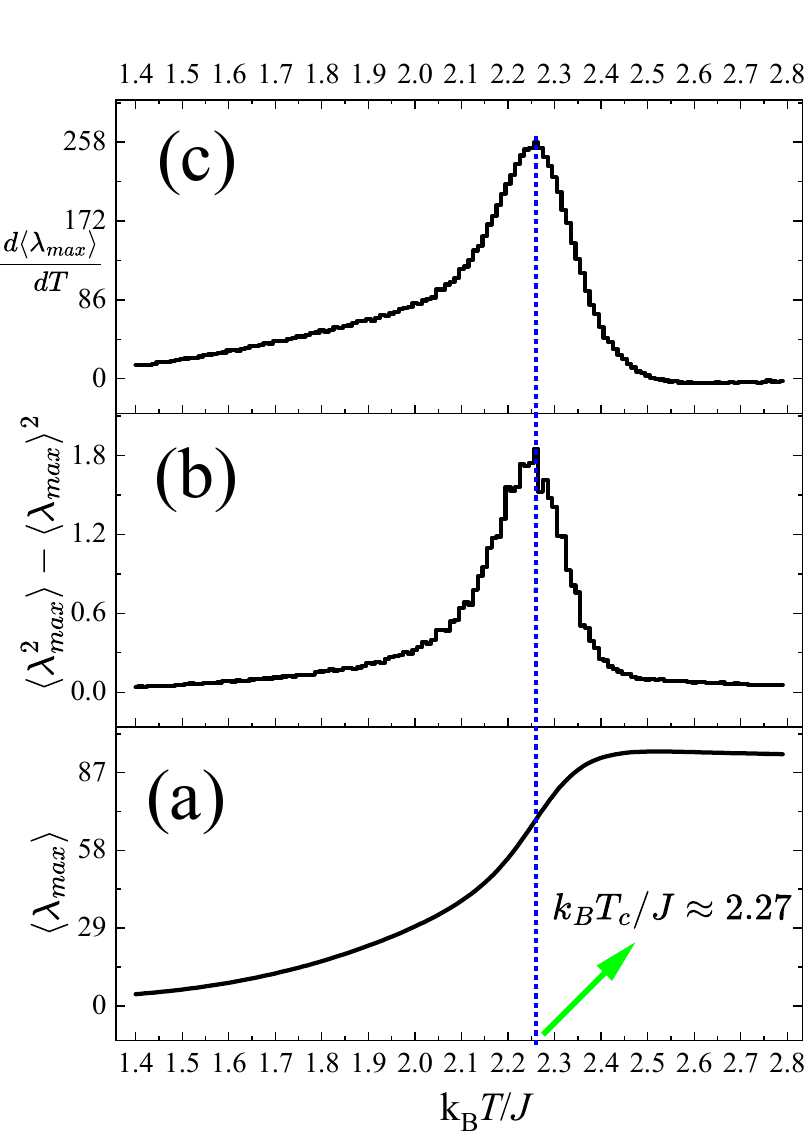}
\end{center}
\caption{Monitoring of $\left\langle \protect\lambda _{\max }\right\rangle $%
, $\left\langle \left( \Delta \protect\lambda \right) ^{2}\right\rangle $,
and $\frac{d\left\langle \protect\lambda _{\max }\right\rangle }{dT}$ as
function of $k_{B}T/J$ shown in panels (a), (b), and (c), respectively. An
inflection point is observed for $\left\langle \protect\lambda _{\max }\right\rangle $ in 
$k_{B}T_{c}/J\approx 2.27$, corresponding to maxima of $\left\langle
\left( \Delta \protect\lambda \right) ^{2}\right\rangle $ and $\frac{%
d\left\langle \protect\lambda _{\max }\right\rangle }{dT}$ illustrating that $%
\protect\lambda _{\max }$ behaves as an empirical order parameter of Ising model, 
It should be noted, however, that the fluctuations and the derivative of the average largest 
eigenvalue develop finite maxima rather than the critical divergences expected in the thermodynamic limit.}
\label{Fig:phase_transition_ising}
\end{figure}

We observe an inflection point in Fig. \ref{Fig:phase_transition_ising} (a)
for $\left\langle \lambda _{\max }\right\rangle $ in $\frac{k_{B}T_{c}}{J}%
\approx 2.27$ that coincides with maxima of $\left\langle \left( \Delta
\lambda \right) ^{2}\right\rangle $ and $\frac{d\left\langle \lambda _{\max
}\right\rangle }{dT}$ shown, respectively, in Fig. \ref%
{Fig:phase_transition_ising} (b) and (c) \cite{Onsager1944}. These results suggest that $%
\lambda _{\max }$ acts, at least empirically, as an order parameter for the Ising model for the transition. Rather than exhibiting a divergence in $\left\langle \left( \Delta \lambda \right)
^{2}\right\rangle $ and $\frac{d\left\langle \lambda _{\max }\right\rangle }{%
dT}$ these quantities display finite maxima that coincide with the inflection
point of $\left\langle\lambda_{\max}\right\rangle$. 

Although the Tracy--Widom distribution constitutes the natural theoretical
description for the largest eigenvalue of pure Wishart ensembles, its
evaluation is computationally demanding, since it is defined in terms of
Fredholm determinants or, equivalently, solutions of the Painlev\'{e} II
differential equation \cite{TracyWidom1994,Bornemann2010}. Therefore, for the purpose of quantitatively
characterizing the departure from Gaussian behavior, it is convenient to
employ a simpler analytical distribution capable of reproducing its main
qualitative features.

A natural choice is the skew-normal distribution \cite{Azzalini1985}, whose probability density
function is given by

\begin{equation}
f(x)=\frac{2}{\omega }\phi \!\left( \frac{x-\xi }{\omega }\right) \Phi
\!\left( \alpha \frac{x-\xi }{\omega }\right) ,  \label{Eq:skew_normal}
\end{equation}%
where

\begin{equation}
\phi (x)=\frac{1}{\sqrt{2\pi }}e^{-x^{2}/2}
\end{equation}%
is the standard normal density and

\begin{equation}
\Phi (x)=\frac{1}{2}\left[ 1+\mathrm{erf}\left( \frac{x}{\sqrt{2}}\right) %
\right]
\end{equation}%
is its cumulative distribution function. The parameters $\xi $, $\omega $
and $\alpha $ denote the location, scale and shape parameters, respectively.
For $\alpha =0$, the skew-normal distribution reduces exactly to the
Gaussian distribution,

\begin{equation}
f(x)=\frac{1}{\omega \sqrt{2\pi }}\exp \left[ -\frac{(x-\xi )^{2}}{2\omega
^{2}}\right] ,  \label{Eq:gaussian}
\end{equation}%
whereas nonzero values of $\alpha $ continuously introduce asymmetry into
the distribution.

In the present work, the skew-normal distribution is not intended to replace
the Tracy--Widom law, but rather to provide a convenient phenomenological
description of asymmetric fluctuations. By comparing Gaussian and
skew-normal fits to the numerical distributions of the largest eigenvalue,
one can quantify the emergence of asymmetry and investigate possible
crossovers between nearly Gaussian and strongly asymmetric regimes.

We therefore perform fits with a simple gaussian function (Eq. \ref%
{Eq:gaussian}) and with a skew-normal distribution given by Eq. \ref%
{Eq:skew_normal}.

\begin{table}[tbph]
\caption{Comparison of Gaussian and skew‑normal (Tracy--Widom–like) fits to the distribution of \(\lambda_{\max}\).}
\label{tab:gauss_tw_comparison}\centering
\resizebox{\textwidth}{!}{\begin{tabular}{c c c c c c c c c c c c c}
\hline
$T$ & $\langle \lambda_{\max}\rangle$ & $\sigma$ & skew. & kurt. & KS$_G$ & $p_G$ & KS$_{TW}$ & $p_{TW}$ & $\Delta KS$ & $R$ & shape$_{TW}$ & best \\
\hline
TC-1.5 & 2.76501 & 0.0841694 & 0.393827 & 0.295771 & 0.0254937 & 4.44363e-06 & 0.00763936 & 0.600986 & 0.0178544 & 0.299657 & 1.75384 & TW \\
TC-1.3 & 2.80732 & 0.0853386 & 0.397371 & 0.290793 & 0.031487 & 4.77247e-09 & 0.00606717 & 0.853005 & 0.0254198 & 0.192688 & 1.794 & TW \\
TC-1.0 & 3.14321 & 0.134787 & 0.359694 & 0.0618349 & 0.0290268 & 9.3984e-08 & 0.00858499 & 0.450015 & 0.0204418 & 0.29576 & 1.77533 & TW \\
TC-0.5 & 15.132 & 0.370442 & 0.0325753 & 0.0261771 & 0.00501127 & 0.962139 & 0.00346781 & 0.999722 & 0.00154346 & 0.692002 & 0.560847 & TW \\
TC-0.05 & 59.3871 & 1.27051 & 0.0466579 & 0.00256237 & 0.00575469 & 0.893047 & 0.00434428 & 0.99125 & 0.00141041 & 0.754911 & 0.641295 & TW \\
TC-0.02 & 66.7265 & 1.32298 & -0.02151 & -0.0438339 & 0.00458325 & 0.984004 & 0.00554008 & 0.917016 & -0.000956829 & 1.20877 & -0.459059 & Gauss \\
TC & 71.7768 & 1.31202 & 0.00586654 & -0.0447847 & 0.00702569 & 0.704187 & 0.00679239 & 0.742756 & 0.000233296 & 0.966794 & 0.267572 & TW \\
TC+0.05 & 83.1098 & 1.04139 & -0.151044 & -0.00608705 & 0.0132341 & 0.0596914 & 0.00741726 & 0.638312 & 0.0058168 & 0.560467 & -1.04614 & TW \\
TC+0.5 & 93.9691 & 0.230692 & -0.146174 & 0.0636238 & 0.0117888 & 0.123134 & 0.00435665 & 0.990949 & 0.00743214 & 0.369559 & -1.04275 & TW \\
TC+1.0 & 92.9054 & 0.154662 & -0.0881468 & 0.0161211 & 0.00923971 & 0.358323 & 0.00384511 & 0.998343 & 0.0053946 & 0.41615 & -0.823401 & TW \\
TC+3.0 & 88.5199 & 0.128334 & 0.0163486 & 0.0583358 & 0.0084776 & 0.466199 & 0.00708197 & 0.694772 & 0.00139563 & 0.835375 & 0.475947 & TW \\
TC+10.0 & 72.3639 & 0.227645 & 0.0352023 & -0.0858476 & 0.00746165 & 0.630831 & 0.00728178 & 0.661161 & 0.000179866 & 0.975895 & 0.54854 & TW \\
\hline
\end{tabular}}
\end{table}

The columns reported in Table~\ref{tab:gauss_tw_comparison} summarize the
statistical properties of these values and compare Gaussian and skew-normal
descriptions of their empirical distribution.

The first column, denoted by $T$, identifies the temperature at which the
ensemble of correlation matrices was generated. The average largest
eigenvalue $\left\langle \lambda _{\max }\right\rangle $ for each temperature is presented in the second column.

The standard deviation, represented by $\sigma$, is calculated according to

\begin{equation}
\sigma =\left[ \frac{1}{N_{\mathrm{run}}}\sum_{r=1}^{N_{\mathrm{run}}}\left(
\lambda _{\max }^{(r)}-\left\langle \lambda _{\max }\right\rangle \right)
^{2}\right] ^{1/2}.  \label{eq:std_largest_eigenvalue}
\end{equation}%
is presented in the third column. We then consider the higher moments. The column denoted by skewness contains the normalized third central moment,

\begin{equation}
\gamma_{1} = \frac{ \displaystyle \frac{1}{N_{\mathrm{run}}} \sum_{r=1}^{N_{%
\mathrm{run}}} \left( \lambda_{\max}^{(r)} -
\left\langle\lambda_{\max}\right\rangle \right)^{3} }{ \sigma^{3} }.
\label{eq:sample_skewness}
\end{equation}

For a symmetric distribution, $\gamma_{1}=0$. Positive values, $\gamma_{1}>0$%
, indicate a longer right tail, associated with unusually large values of $%
\lambda_{\max}$, whereas negative values indicate a longer left tail.
Therefore, the skewness provides a direct measure of the departure from
symmetry.

The kurtosis column reports the excess kurtosis,

\begin{equation}
\gamma_{2} = \frac{ \displaystyle \frac{1}{N_{\mathrm{run}}} \sum_{r=1}^{N_{%
\mathrm{run}}} \left( \lambda_{\max}^{(r)} -
\left\langle\lambda_{\max}\right\rangle \right)^{4} }{ \sigma^{4} } -3.
\label{eq:sample_excess_kurtosis}
\end{equation}

The subtraction of $3$ is introduced so that a Gaussian distribution has $%
\gamma _{2}=0$. Positive values of $\gamma _{2}$ indicate a more pronounced
central peak and/or heavier tails than those of a Gaussian distribution.
Negative values are associated with a flatter distribution and/or lighter
tails.

The quality of each fit is quantified through the Kolmogorov--Smirnov
statistic \cite{Durbin1973}. For the Gaussian fit, it is defined as

\begin{equation}
D_{\mathrm{G}}=\sup_{x}\left\vert F_{\mathrm{emp}}(x)-F_{\mathrm{G}%
}(x)\right\vert ,  \label{eq:ks_gaussian}
\end{equation}%
where $F_{\mathrm{emp}}(x)$ is the empirical cumulative distribution
function and $F_{\mathrm{G}}(x)$ is the cumulative distribution function of
the fitted Gaussian model. The corresponding quantity for the skew-normal
fit is

\begin{equation}
D_{\mathrm{TW}}=\sup_{x}\left\vert F_{\mathrm{emp}}(x)-F_{\mathrm{TW}%
}(x)\right\vert .  \label{eq:ks_skew_normal}
\end{equation}

These quantities are reported in the table as $\mathrm{KS}_{\mathrm{G}}$ and 
$\mathrm{KS}_{\mathrm{TW}}$. In both cases, smaller values indicate that the
fitted cumulative distribution is closer to the empirical one. For example,
a value $\mathrm{KS}_{\mathrm{G}}=0.01$ means that the maximum vertical
distance between the empirical and Gaussian cumulative distributions is
approximately one percent.

The columns $p_{\mathrm{G}}$ and $p_{\mathrm{TW}}$ contain the nominal $p$%
-values returned by the Kolmogorov--Smirnov test. In the conventional
interpretation, a small $p$-value indicates evidence against the
corresponding model. However, in the present analysis, the distribution
parameters are estimated using the same data employed in the test.
Therefore, the standard Kolmogorov--Smirnov $p$-values are not exact and
should be interpreted only as approximate indicators \cite{Lilliefors1967,Durbin1973}. A formally calibrated
significance analysis would require a parametric bootstrap procedure \cite{Stute1993}.

To compare the distances obtained for the two models, the difference

\begin{equation}
\Delta \mathrm{KS}=D_{\mathrm{G}}-D_{\mathrm{TW}}  \label{eq:delta_ks}
\end{equation}%
is also reported. Thus,

\begin{equation}
\Delta \mathrm{KS}>0
\end{equation}%
indicates that the skew-normal distribution has a smaller
Kolmogorov--Smirnov distance, while

\begin{equation}
\Delta \mathrm{KS}<0
\end{equation}%
indicates that the Gaussian distribution provides the smaller distance.
Values close to zero imply that both descriptions have essentially the same
performance according to this criterion.

The ratio

\begin{equation}
R=\frac{D_{\mathrm{TW}}}{D_{\mathrm{G}}}  \label{eq:ks_ratio}
\end{equation}%
provides a normalized measure of the relative quality of the two fits. Its
interpretation is

\begin{equation}
R<1\quad \Longrightarrow \quad D_{\mathrm{TW}}<D_{\mathrm{G}},
\label{eq:ratio_sn_better}
\end{equation}%
and therefore the skew-normal fit is closer to the empirical distribution.
Conversely,

\begin{equation}
R>1\quad \Longrightarrow \quad D_{\mathrm{G}}<D_{\mathrm{TW}},
\label{eq:ratio_gaussian_better}
\end{equation}%
so that the Gaussian fit is preferred according to the Kolmogorov--Smirnov
distance. A value $R\simeq 1$ indicates that both fits describe the data
with comparable accuracy. For instance, $R=0.25$ means that the skew-normal
Kolmogorov--Smirnov distance is only $25\%$ of the Gaussian one, whereas $%
R=0.98$ indicates only a marginal improvement. The second-to-last column presents the \textrm{shape}$_{TW}$ which corresponds to the parameter $\alpha$ in Eq. \ref{Eq:skew_normal}.

Finally, the column denoted by ``best'' identifies the distribution with the
smallest Kolmogorov--Smirnov statistic,

\begin{equation}
\mathrm{best} = \left\{ 
\begin{array}{ll}
\mathrm{Skew\mbox{-}normal}, & D_{\mathrm{SN}}<D_{\mathrm{G}}, \\[4pt] 
\mathrm{Gaussian}, & D_{\mathrm{G}}\leq D_{\mathrm{SN}}.%
\end{array}
\right.  \label{eq:best_fit_definition}
\end{equation}

\begin{figure}[tbp]
\begin{center}
\includegraphics[width=1.0\columnwidth]{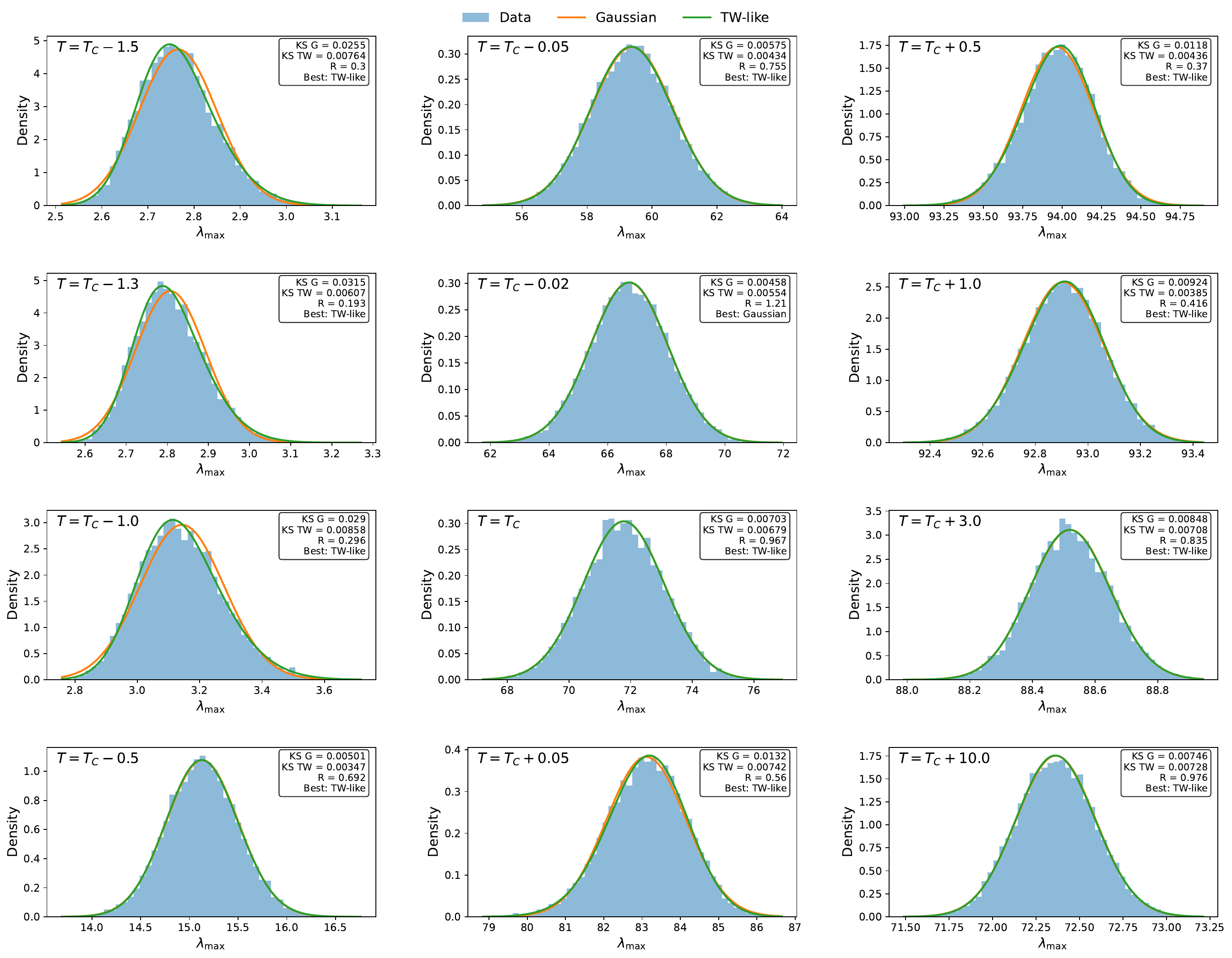}
\end{center}
\caption{Gaussian and skew-normal fits for the different temperatures. }
\label{Fig:fits}
\end{figure}

These results show that the approach to criticality is accompanied by
changes in the shape of the largest-eigenvalue distribution. A nearly
Gaussian behavior is found slightly below the critical temperature, at
$T=T_{c}-0.02$, and occurs only within a narrow temperature interval in the
present data. The fits obtained at the different temperatures are shown in
Fig. \ref{Fig:fits}, and the distribution shapes corroborate the numerical
comparisons reported in Table \ref{tab:gauss_tw_comparison}.

To further test the idea of characterizing transitions through
$\lambda _{\max }$ and its distribution, we repeat the procedure for the
transition to chaos. We iterate the logistic map \cite{May1976} and construct ensembles of correlation matrices for different values of $r$. For this analysis, we again use $N_{run}=1000$, $T=300$, and $N=100$.

\begin{figure}[tbp]
\begin{center}
\includegraphics[width=0.9\columnwidth]{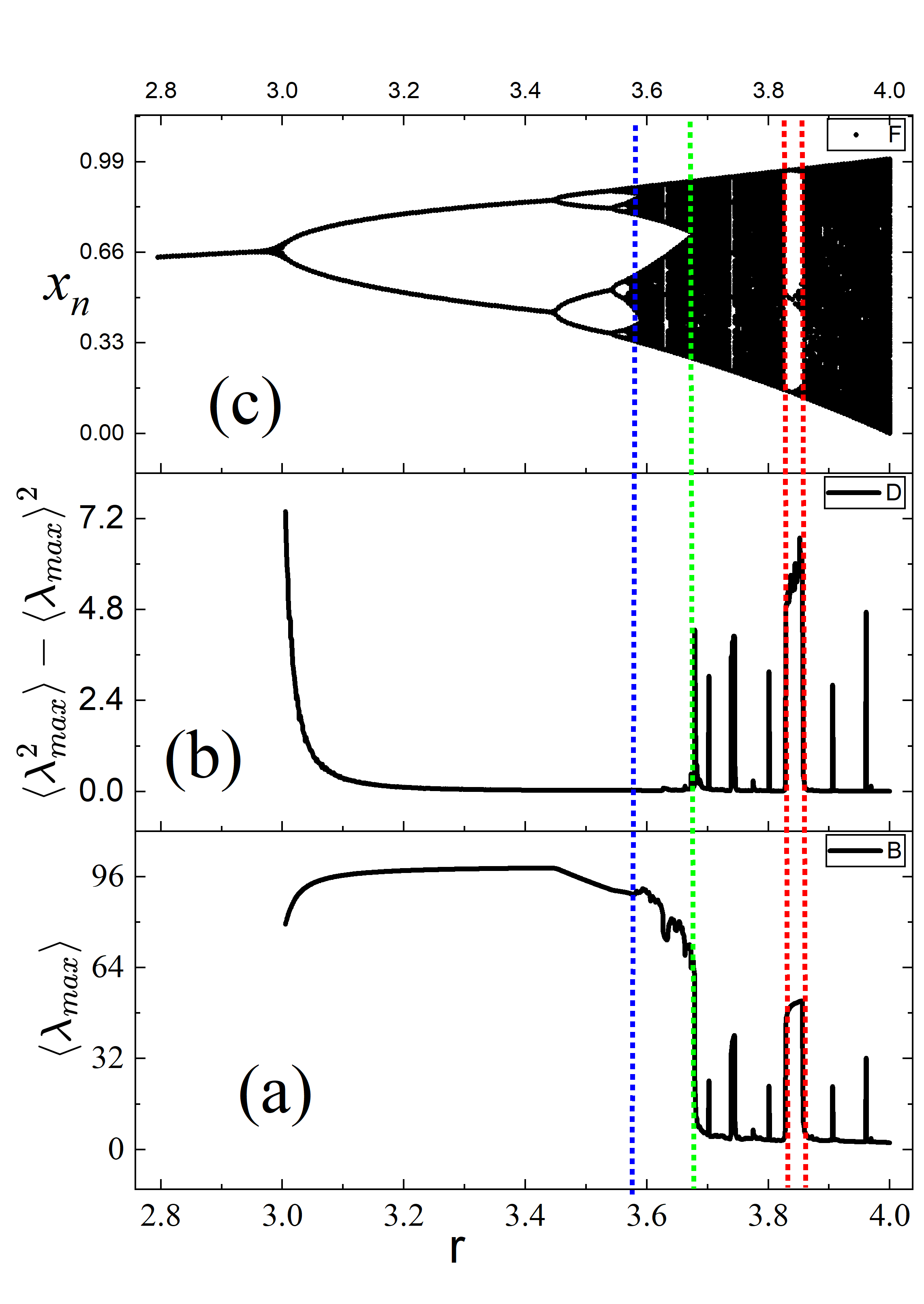}
\end{center}
\caption{Spectral characterization of the transition to chaos using $\protect\lambda _{\max }$. Panel (a) shows the behavior of $\left\langle \protect\lambda _{\max }\right\rangle $
as a function of $r$. A sharp variation occurs for $r\approx 3.67$. This value corresponds to a maximum in the variance of the largest
eigenvalue, shown in panel (b). Panel (c) shows the corresponding features in the bifurcation
diagram. }
\label{Fig:transition_to_chaos}
\end{figure}

Figure \ref{Fig:transition_to_chaos}(a) shows the behavior of $\left\langle
\lambda _{\max }\right\rangle $ as a function of $r$. A sharp variation occurs
for $r\approx 3.67$, in contrast with the Ising case, in which the transition is associated with
an inflection point. This value also corresponds to a maximum in the variance of the largest
eigenvalue, shown in panel (b), demonstrating that $\lambda_{\max}$ again
carries a clear signature of the transition, in this case the transition to
chaos. It is also important to compare these results with the bifurcation diagram
shown in panel (c).
The blue line corresponds to the critical point of Feigenbaum $r_{c}\approx
3.57$ which marks the accumulation point of the period-doubling cascade and the
conventional onset of chaos \cite{Feigenbaum1978}. In $\left\langle \lambda _{\max}\right\rangle$, this point appears as a
small shoulder. A more pronounced spectral change occurs near $r\approx3.66$, where the
bifurcation diagram develops the dense dark region highlighted by the green
line. This feature coincides with the sharp change in
$\left\langle\lambda_{\max}\right\rangle$ identified by our construction.
Thus, the spectral criterion emphasizes a later stage in the development of
chaotic dynamics than the Feigenbaum accumulation point itself. 

The white gaps correspond to periodic windows of different periods \cite{May1976}, which
produce spikes in $\left\langle \lambda _{\max }\right\rangle $ and
in $\left\langle \lambda _{\max }^{2}\right\rangle -$ $\left\langle \lambda
_{\max }\right\rangle ^{2}$. It is important to highlight the periodic window
of period 3 that starts in $r_{c}^{(1)}\approx 3.82$ and finishes in $%
r_{c}^{(2)}\approx 3.85$ \cite{LiYorke1975}, which is indicated by the two red lines. 

\begin{figure}[tbp]
\begin{center}
\includegraphics[width=1.0\columnwidth]{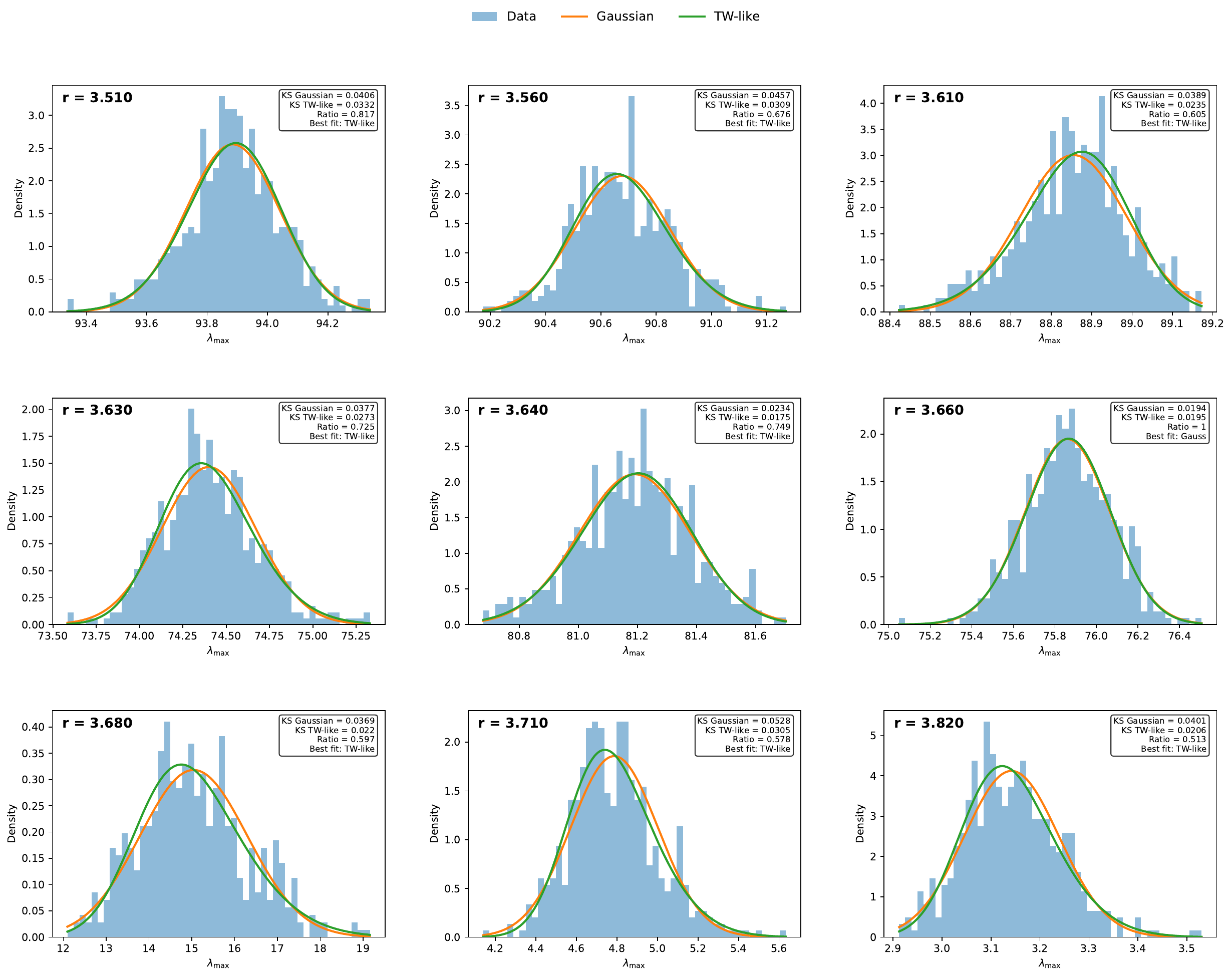}
\end{center}
\caption{Gaussian and skew-normal fits for the different values of $r$.}
\label{Fig:logistic_histograms}
\end{figure}

Finally, we also examine the histograms of $\lambda
_{\max }$ as done for the Ising model. The results reveal an interesting pattern as shown in Fig. \ref{Fig:logistic_histograms}. 

Near $r\approx3.66$, the skew-normal (orange) and Gaussian (green) fits
become almost indistinguishable. In this narrow region, the Gaussian fit is
slightly favored, coinciding with the sharp variation of
$\left\langle\lambda_{\max}\right\rangle$. The legends report the
corresponding $\mathrm{KS}$ statistics and the ratio $R$. In particular,
$R\simeq1$ near $r\approx3.66$, indicating that both distributions provide
comparably accurate descriptions precisely where the spectral observable
changes most sharply.  

\section{Summaries and Conclusions}

\label{Sec:Conclusions}

In this work, we investigated the largest eigenvalue of correlation matrices
as a compact spectral observable for identifying changes in collective
organization. The central motivation is that
$\lambda_{\max}$ measures the maximum variance carried by a normalized
collective mode. Consequently, it is particularly sensitive to correlations
that are coherently distributed among many variables. Within the Wishart
framework, the Marchenko--Pastur law and the statistics of the spectral edge
provide a natural random reference against which this collective component
can be assessed.

For the two-dimensional Ising model, the average largest eigenvalue exhibits
a clear inflection point near the exact critical temperature,
$k_{B}T_{c}/J\simeq2.27$. The same temperature is identified by maxima in the
variance of $\lambda_{\max}$ and in the derivative
$d\langle\lambda_{\max}\rangle/dT$. These results show that the largest
eigenvalue responds directly to the reorganization of correlations at the
phase transition. Although $\lambda_{\max}$ is not a thermodynamic order
parameter in the strict sense, its mean value and fluctuations provide an
effective empirical spectral indicator of criticality.

The analysis of the full distribution of $\lambda_{\max}$ supplies
additional information that is not contained in its mean value alone. The
comparison between Gaussian and skew-normal fits reveals temperature-dependent
changes in asymmetry and tail structure. In particular, a nearly Gaussian
distribution is observed in a narrow region slightly below the critical
temperature, whereas asymmetric distributions are favored over most of the
investigated range. The skew-normal law is used here only as a convenient
phenomenological proxy for asymmetric edge fluctuations and should not be
identified with the exact Tracy--Widom distribution. Nevertheless, the
crossover between symmetric and asymmetric shapes indicates that the
distribution of the extreme eigenvalue contains complementary information
about the restructuring of collective fluctuations.

The same methodology was applied to the logistic map. The average largest
eigenvalue, its variance, and the distributional fits reproduce several
features of the bifurcation diagram, including periodic windows and the
development of strongly chaotic regions. The conventional Feigenbaum
accumulation point near $r\simeq3.57$ appears as a shoulder in the spectral
observable, while a sharper change occurs near $r\simeq3.66$, where the
bifurcation diagram develops a dense chaotic band. Therefore, the proposed
spectral criterion does not simply reproduce the Lyapunov-exponent
definition of the onset of chaos. Instead, it identifies a pronounced
reorganization of the ensemble correlation structure generated by the map.

Taken together, the two applications demonstrate that the largest eigenvalue
can serve as a model-independent detector of structural changes in systems
whose microscopic dynamics are very different. The Ising model undergoes an
equilibrium phase transition driven by collective spin fluctuations, whereas
the logistic map develops chaos through a sequence of dynamical
bifurcations. In both cases, these changes leave measurable fingerprints in
$\langle\lambda_{\max}\rangle$, in its variance, and in the shape of its
probability distribution.

Several limitations should be emphasized. The locations and sharpness of the
spectral features depend on the matrix dimension, the length of the sampled
time series, the number of realizations, the initial conditions, and the
specific observable used to construct the matrices. A systematic finite-size
and finite-time scaling analysis is therefore required before universal
critical exponents or asymptotic laws can be inferred. In addition, the
nominal Kolmogorov--Smirnov $p$-values reported here are only approximate
because the fitting parameters are estimated from the same samples used in
the tests. Parametric-bootstrap calibration and a direct numerical
comparison with the Tracy--Widom law would provide a more rigorous
distributional assessment.

Despite these limitations, the present results support the largest
eigenvalue of correlation matrices as a useful empirical order parameter and
as a practical diagnostic of collective reorganization. The method requires
only ensembles of time evolutions and does not depend on prior knowledge of a
conventional order parameter. This feature makes it potentially valuable for
complex systems in which the relevant macroscopic variable is unknown,
difficult to measure, or absent in the traditional thermodynamic sense. 

Methods based on the analysis of time series can also characterize critical 
phenomena by considering the entire eigenvalue spectrum through the moments of 
the eigenvalue density, as demonstrated in Ref.~\cite{daSilva2023}. Although 
this approach provides a more complete spectral description, it is computationally 
more demanding. Alternatively, time series can be mapped into visibility graphs \cite{Lacasa2008}, 
whose Laplacian spectra also reveal signatures of criticality \cite{daSilva2026}. In 
that case, however, the analysis requires the computation of determinants associated 
with the number of spanning trees. In contrast, the present work shows that the largest 
eigenvalue alone is sufficient to characterize the critical behavior, leading to a considerably 
simpler and computationally more efficient approach.  

\bigskip

\bibliographystyle{unsrt}
\bibliography{largest_eigenvalue_references}

\end{document}